\documentclass{kapproc} 
\let\footnote\savefootnote
\let\footnotetext\savefootnotetext 
\let\footnoterule\savefootnoterule 
\kluwerbib

\RequirePackage{graphicx}%
\RequirePackage{epsf}%

\newcommand{\HIbold}{{H\footnotesize{\bf{I}}}}
\newcommand{\HI}{\mbox{\sc H{i}}}

\newcommand{\msun}{\mbox{$M_\odot$}}

\newcommand{\kms}{\mbox{km s$^{-1}$}}
\newcommand{\cm}{{cm}$^{-2}$}
\newcommand{\Lsun}{\mbox{$L_\odot$}}
\newcommand{\ha}{\mbox{H$\alpha$}}
\newcommand{\mhilb}{\mbox{$M_{\rm HI}$-to-$L_B$}}

\begin{document}

\articletitle{Galactic Recycling: \\ The \HIbold\ Ring Around NGC 1533.}

\author{Emma Ryan-Weber \& Rachel Webster}
\affil{University of Melbourne}
\email{eryan@physics.unimelb.edu.au, rwebster@physics.unimelb.edu.au}
\author{Kenji Bekki}
\affil{University of New South Wales}
\email{bekki@bat.phys.unsw.edu.au}

\chaptitlerunninghead{\HI\ Ring Around NGC 1533}

\begin{abstract}
We report the discovery of a new \HI\ ring around the S0 galaxy NGC
1533. The ring orbits at a radius of 35 kpc, well outside the optical
extent of the galaxy. We have conducted N-body/SPH numerical simulations to
show this \HI\ ring could be the merger remnant of a tidally destroyed
galaxy. We find no optical component associated with the \HI\
ring. However, observations hint at \ha\ emission associated with
the SE part of the ring only. The \ha\ is in the form
of a few very small isolated emission line regions. The large \HI\
velocity dispersions (up to 30 \kms) and velocity gradients (up to 50
\kms kpc$^{-1}$) in this region indicate the \ha\ emission could be
due to star formation triggered by clouds colliding within the ring.
\end{abstract}

\section{Introduction}
In the local Universe galaxies continue to interact and merge. These
mergers provide feedback into the intergalactic medium; either
directly or via star formation. S0 galaxies are synonymous with
non-regular \HI\ distributions - leading to the common perception that
this gas was acquired via accretion or mergers. Simple passive
evolution of elliptical and S0 galaxies is inconsistent with
observations. To explain the observations, merging and star formation
in these galaxy types must have occurred from $z\sim1$ to the present
(Kauffmann, Charlot \& White 1996). In contrast to early-type spirals,
S0 galaxies exhibit a wide range of \mhilb\ ratios. Wardle \& Knapp
(1986) argue this is evidence for an external origin of \HI\ in
S0s. This acquisition of \HI\ is perhaps more notable in S0s, since
their intrinsic \HI\ content is low. Similar events in spiral galaxies
may not be detected as easily.

The formation of \HI\ rings are rare. Rings are known around
individual galaxies such the spiral galaxy NGC 628 (Briggs, 1982) and
the elliptical galaxy IC 2006 (Schweizer, van Gorkom \& Seitzer
1989). \HI\ rings are also found in galaxy groups enclosing more than
one galaxy, for example, the M96 group (Schneider, 1985) and the
galaxy group LGG 138 (Barnes, 1999). The Cartwheel galaxy, in the
\HI-rich Cartwheel group, also exhibits an \HI\ ring (Higdon 1996).

Explanations for each of these rings invariably involves some form of
merging or accretion. The origin of NGC 628's \HI\ ring is uncertain,
but the absence of a massive companion points toward the acquisition
of a gas-rich dwarf galaxy. The \HI\ ring around IC 2006 is thought to
be the remnant of the merger that created the elliptical or perhaps a
later accretion event. Barnes (1999) proposed the \HI\ ring in LGG 138
was created by a gas-sweeping collision between one of two bright
galaxies and an intruder. Analysis of the ring in the M96
group is complicated by the number of galaxies in the vicinity. The
distribution of \HI\ in M96 itself suggests it is interacting with
the ring and perhaps accreting \HI\ onto its own faint optical outer
ring. The Cartwheel galaxy is believed to have formed from a small
late type spiral with a large low surface density gas disk. Higdon
(1996) suggests that another member of the group, G3, passed through
this disk and `splashed-out' the \HI\ to form the ring.

Optical counterparts to these \HI\ rings are rarer still. A very faint
dwarf galaxy, Leo dw A resides in the M96 ring (Schneider, 1989). The
\HI\ in LGG 138 aligns with a colour break in stellar populations in
the South-western region of the ring. Barnes (1999) suggests this is
due to star formation triggered by an expanding density wave, together
with the stellar remnant of the intruder. On the other hand, the
Cartwheel galaxy was first noted for its remarkable optical ring and
the \HI\ observations followed. In this case, the `splashed-out' \HI\
is thought to have caused a propagating burst of massive star
formation (Higdon, 1996).

When galaxies collide during a merger, they can produce strong shocks,
and stars may then form in the cool, compressed gas behind the shock
front. For example, large \HI\ and CO velocity dispersion and
gradients, in the youngest star forming regions in the Antennae
galaxies, indicate stars were produced by colliding gas clouds (Zhang,
2001). Alternatively, gravitational instabilities can cause collapse
and formation of stars. The balance between self gravity, velocity
dispersion and the centrifugal force in a disk, leads to a critical
surface density, $\Sigma_{crit}= \kappa\sigma_v/\pi G$, where $\kappa$
is the epicycle frequency and $\sigma_v$ is the velocity dispersion of
the gas. According to the Toomre criterion (Toomre, 1964), large scale
star formation occurs when $Q$ $(=\Sigma_{crit}/\Sigma_{gas}$) is less
than one. The \HI\ ring in the Cartwheel galaxy satisfies this
criterion (Higdon, 1996).

Here we present the \HI\ ring surrounding the S0 galaxy NGC 1533. The
ring was discovered serendipitously as part of a subset of galaxies
from HIPASS (see e.g. Meyer et al., 2002) chosen for mapping with the
Australia Telescope Compact Array (ATCA)\footnote{The Australia
Telescope Compact Array is part of the Australia Telescope which is
funded by the Commonwealth of Australia for operation as a National
Facility managed by CSIRO.}. NGC 1533 is located 1$^{\circ}$ from the
centre of the Dorado group. Throughout this paper we assume a distance
to NGC 1533 of 21$\pm$4 Mpc (Tonry et al., 2001).

\section{Observations}

NGC 1533 (and consequently its immediate environment) was imaged in
21-cm with the ATCA in three array configurations: the 375, 750D and
1.5D arrays. The dataset was reduced in MIRIAD, it has a restored
beam of $68\arcsec\times65\arcsec$ and a velocity resolution of
3.3~\kms. The RMS noise in the final cube is 3.7 mJy beam$^{-1}$,
corresponding to a 3$\sigma$ column density limit, over a typical line
width of 40~\kms, of 3.2$\times10^{19}$~\cm.

The \HI\ column density map is overlaid on the DSS image of NGC 1533
in Figure 1a. The \HI\ contours increase linearly from 1.6 to 4.0
$\times10^{20}$~\cm. The two smaller galaxies in the NW
corner of the image are IC 2039 (closest to NGC 1533) \& IC 2038. The
higher resolution ATCA image (1.5D array), not included here, clearly
shows \HI\ associated with IC 2038, but not IC 2039. The \HI\ ring
around NGC 1533 consists of two major components, the NW cloud and the
SE cloud. \HI\ Gas with column densities below the lowest contour
close the ring. No obvious optical counterpart to this ring is
seen. The total \HI\ mass of the system (based on the total flux
density from HIPASS of 67.6 Jy beam$^{-1}$\kms) is
7$\times10^{9}$~\msun.  At its minimum and maximum extent, the radius
of \HI\ ring is 2\arcmin\ and 11.7\arcmin\ from the optical centre of
NGC 1533, corresponding to a projected physical length between 12 and
70 kpc.

Observations of \ha\ emission in NGC 1533 and surrounds were taken as
part of the Survey for Ionization in Neutral-Gas Galaxies
(SINGG). Continuum R-band and narrow band \ha\ images were taken with
the CTIO 1.5m telescope. Only one moderate \ha\ emission line region
is seen within the disk of NGC 1533. We also find 5 very small
isolated emission line regions in the SE part of the \HI\ ring (see
Figure 1b). These emission line regions have \ha\ fluxes of
1-2$\times10^{15}$ erg s$^{-1}$\cm. At 21 Mpc, this corresponds to
\ha\ luminosities of 5-10$\times10^{37}$ erg s$^{-1}$.

\section{Analysis}
Both internal and external origins of the \HI\ ring are considered.
If the \HI\ was intrinsic to NGC 1533, it could have been removed via
ram pressure stripping by a denser intragroup gas. In this case NGC
1533 would have had a \mhilb\ ratio of 1.6, which is plausible for an
early-type spiral galaxy. However there is no evidence for a dense
intragroup medium in the Dorado group. Also, simulations do not
produce the ring-like structure we see.

Alternatively, the \HI\ could have been accreted from another galaxy.
IC 2038 and IC 2039 are obvious suspects. The luminosity of these
galaxies are 1.5 and 2.6$\times10^{8}$~\Lsun\ respectively. To account
for all the \HI\ in the system, these galaxies would had to have
had \mhilb\ ratios greater than 15, which is not very likely for their
morphologies.

The third possibility is the tidal destruction of a galaxy to form a
merger remnant around NGC 1533. N-body/SPH numerical simulations were
conducted to investigate the orbital evolution of a low surface
brightness in NGC 1533's gravitational potential. The simulated
galaxies are based on the Fall-Efstathiou (1980) model. Figure 1c
shows the dynamical evolution of the gas in a galaxy approaching on a
highly eccentric orbit. The frame is centred on a static potential for
NGC 1533. The \HI\ ring forms after 2$\times10^{9}$ years (T=14) and
continues to orbit. The ring-like distribution of gas is also
recovered for different orbital entry points and eccentricities. The
surface brightness of the stellar remnant is reduced to 26-29 mag
arcsec$^{-2}$, which explains why we do not see an optical
counterpart. This scenario seems the most plausible explanation for
the \HI\ ring. The simulation also lends support to the merger
hypothesis for other galaxies with \HI\ rings.

Does this recycled \HI\ gas then form stars? The small \ha\ emission line
regions we see in the SE part of the ring are similar to those
reported by Ferguson et al. (1998) in the extreme outer region of disk
galaxies. Two star formation scenarios are investigated below.

Firstly, gravitational instabilities in the \HI\ ring could allow
stars to form. As described in the introduction, large scale star
formation can occur in a disk when $Q<1$. $Q$ was calculated for every
pixel in the 21-cm image. To do this, a tilted ring with p.a.=140\deg\
and {\it{i}}=70\deg\ was fitted to the \HI\ data and a rotation curve
derived. The rotation curve was used to calculate each value of
$\kappa$. The velocity dispersion measurement used spectra clipped at
3$\sigma$ to reduce the effect of noise. Not a single pixel was found
to satisfy the $Q<1$ criterion. $Q$ varies from 2 in the densest part
of the NW cloud, to greater than 10 in the SE cloud. The large values
of $Q$ in the SE region is due to the high velocity dispersions. This
result agrees with the absence of massive star formation in the \HI\
ring, however it fails to explain the \ha\ emission line regions we do
see.

The second possibility is clouds colliding within the ring, condensing
the gas and forming stars. The kinematics of the \HI\ local to each
\ha\ emission line region were analysed. \HI\ velocity dispersions
were found up to 30~\kms\ and velocity gradients in the range
7-50~\kms~kpc$^{-1}$. The \HI\ velocity profiles of some individual
pixels in these region are also `double-horned', indicating expanding
or contracting regions of gas. This is in contrast to the \HI\
kinematics in the NW cloud, there the dispersions are mostly less
than 10~\kms\ and gradients around 2~\kms~kpc$^{-1}$.  These large
gas dispersion and gradients are similar to other merger-driver star
formation examples such as the young star clusters in the Antennae
galaxies.

We conclude that the \HI\ ring around NGC 1533 is most likely the
remnant of a tidally destroyed galaxy. Nbody/SPH numerical simulations
support this hypothesis. \ha\ observations show low level star
formation in the SE part of the ring. The high \HI\ velocity
dispersion and gradients in this region indicate stars may have
formed by clouds colliding in the \HI\ ring that is yet to stabilise.

\section{Acknowledgements}
Thanks to the SINGG group for the \ha\ images. ERW acknowledges the
conference organisers for financial assistance and ATNF, CSIRO
for the student overseas travel award.

\begin{figure}
\plotone{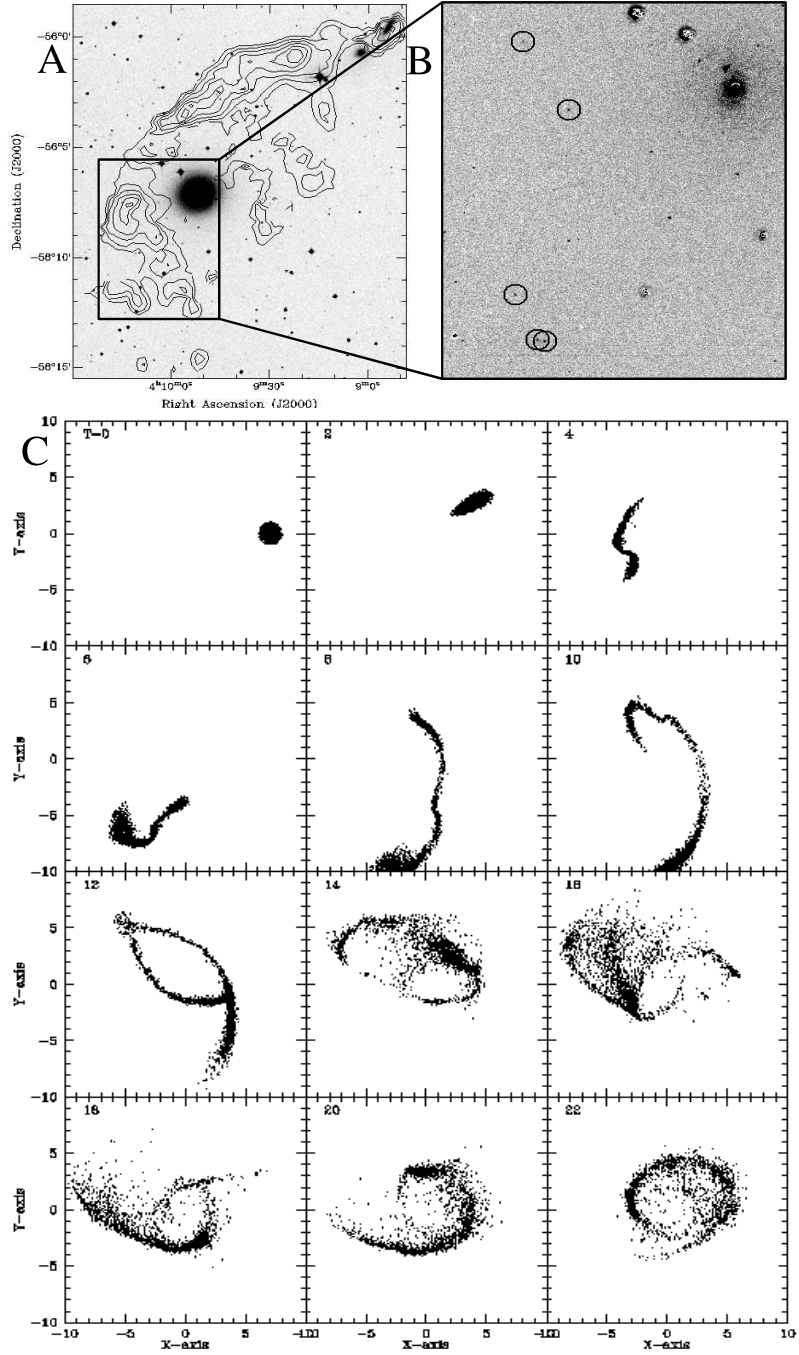}
\label{fig:fig1}
\caption{a) \HI\ ring around NGC 1533. b) SE ring region with 
\ha\ emission line regions circled. c) Gaseous components in the simulation.}
\end{figure}

\end{document}